# A Systematic Approach for Exploring Tradeoffs in Predictive HVAC Control Systems for Buildings

Joshua Gluck   Christian Koehler   Jennifer Mankoff   Anind Dey   Yuvraj Agarwal

Carnegie Mellon University

{jgluck, ckoehler, jmankoff, anind, yuvraj.agarwal}@andrew.cmu.edu

## ABSTRACT

Heating, Ventilation, and Cooling (HVAC) systems are often the most significant contributor to the energy usage, and the operational cost, of large office buildings. Therefore, to understand the various factors affecting the energy usage, and to optimize the operational efficiency of building HVAC systems, energy analysts and architects often create simulations (*e.g.*, EnergyPlus or DOE-2), of buildings prior to construction or renovation to determine energy savings and quantify the Return-on-Investment (ROI). While useful, these simulations usually use static HVAC control strategies such as lowering room temperature at night, or reactive control based on simulated room occupancy. Recently, advances have been made in HVAC control algorithms that *predict room occupancy*. However, these algorithms depend on costly sensor installations and the tradeoffs between predictive accuracy, energy savings, comfort and expenses are not well understood. Current simulation frameworks do not support easy analysis of these tradeoffs. Our contribution is a simulation framework that can be used to explore this design space by generating objective estimates of the energy savings and occupant comfort for different levels of HVAC prediction and control performance. We validate our framework on a real-world occupancy dataset spanning 6 months for 235 rooms in a large university office building. Using the gold standard of energy use modeling and simulation (Revit and Energy Plus), we compare the energy consumption and occupant comfort in 29 independent simulations that explore our parameter space. Our results highlight a number of potentially useful tradeoffs with respect to energy savings, comfort, and algorithmic performance among predictive, reactive, and static schedules, for a stakeholder of our building

## Categories and Subject Descriptors

[**Computer systems organization**]: Special purpose systems, sensors and actuators, real-time system architecture

## 1. INTRODUCTION

According to the U.S. Department of Energy, buildings constitute about 41% of primary energy usage in the U.S., with commercial buildings constituting half of that. Heating, Ventilation, and Cooling (HVAC) systems in commercial buildings account for about 40% of this energy use within buildings, contributing to high operational costs [34]. Given this significant expense, buildings are generally first created as software representations (e.g. building models), and a number of tools have been developed to facilitate efficient design and renovation of buildings, such as EnergyPlus [12], DOE-2 [8], and Design-Builder [32]. These tools are used for a variety of purposes, including verifying design functionality, and determining Return-on-Investment (ROI) for building design elements, such as various forms of insulation and HVAC components and control systems [24][28][30][37]. There have been a number of further extensions of these tools in recent years, such as MLE+ [7] and BCTVB [36], which have added functionality, such as making the analysis of output easier or incorporating more complex control strategies [19].

HVAC control systems have advanced along with building simulation tools have advanced. While many modern buildings still use static schedules to run HVAC systems, thereby wasting energy when spaces are unoccupied [5][15][16][17][35][38], recent work has focused on using near real-time occupancy information [5][15], and/or predicted occupancy-based on learned patterns over time [23][31]. Depending on the system, estimated HVAC energy savings could be as high as 30-40% [15][31].

Occupancy-based HVAC actuation substantially reduces the energy usage of buildings while maintaining occupant comfort. However, as evidenced by the fact that most buildings still use static schedules, there are tradeoffs in using occupancy-based HVAC actuation, most notably in the costs of design, installation, and maintenance of an occupancy data detection and inference network. Erickson *et al*. [15] report an expense of $140,000 just for the hardware to support occupancy-based HVAC actuation of a three-floor building, and even simple wireless detection sensors can cost well into the hundreds of thousands of dollars for entire buildings. Research has shown that the energy savings for certain real-time occupancy based control systems [5][15], as well as predictive occupancy based systems [23][31], can make these costs worthwhile, with ROI time being as little as a year. Unfortunately, these studies only explore the ROI for their specific system and use case. So while such studies demonstrate that there are certainly savings to be had for some buildings, they do not provide a methodical way to determine the optimal system among several choices for a given building, as they provide little guidance for exploring the tradeoff-space of energy usage, occupant comfort, cost, and algorithmic performance.

Our goal is to support the incorporation of an occupancy detection and inference system into the simulations systems currently used. This will allow energy analysts, architects, and building managers (*e.g.*, building stakeholders) to explore the tradeoffs between energy usage and occupant comfort at various costs and algorithmic performances of candidate HVAC control systems. Our simulation framework explores the effect of varying parameters of occupancy detection and inference and HVAC control algorithms on energy consumption and occupant comfort. Those parameters include: reference occupancy patterns; false positive and false negative rates of the prediction algorithm; prediction look ahead length; prediction error length; temperature setback settings. *Prediction look ahead length* is defined as how far in advance occupancy predictions occur, *prediction error length* is the temporal clustering of consecutive predictions, and *temperature setback settings* represent the minimum and maximum temperatures that the indoor temperature can reach before the HVAC system engages. A building stakeholder could vary the building model and factors associated with the building (*e.g.*, HVAC system installed, insulation, number and placement of windows) as is currently done by energy analysts and architects when designing or renovating buildings.

A building stakeholder using our framework can input the reported occupancy detection and inference systems' performance of one or more HVAC control systems under consideration, take the output of our simulation and combine it with existing building modeling software (*e.g.*, Energy Plus) and receive objective estimates that support comparison of the amount of energy consumed and occupant discomfort caused by that system. These estimates, in combination with installation and upkeep cost estimates provided by system manufacturers, will allow for a more informed ROI analysis, leading to greater transparency and overall efficiency.

We demonstrate and validate the utility of our framework for assessing energy consumption and occupant comfort using a 196-day longitudinal occupancy-temperature dataset that we have collected for 235 offices in an actual office building. We analyze the effects of varying three parameters in a simulated prediction and control system: *false positive rate* (false positives/(false positives + true negatives)), *false negative rate* (false negatives/(false negatives + true positives)), and *temperature bounds*. We use a Revit building model [4] of a large university office building at CMU and Energy Plus [12] to simulate HVAC control in response to predicted occupancy and weather data. We compare the relative benefits of different levels of false positive and false negative rates, and temperature bounds based on energy consumption, as reported by Energy Plus, and occupant comfort, based on MissTime [25], the average daily number of minutes a room was not at a chosen comfort temperature. This metric indicates on average how many minutes per day an occupant had to endure temperatures that were not their preferred temperatures.

From this demonstration data we were able to derive a number of interesting results. First, in this specific scenario, a predictive occupancy system with medium sized temperature bounds uses 20.8% less energy than a reactive system with small temperature bounds, and 17% less energy than a static schedule system, with comparable occupant comfort for our building. For this scenario, we also show that differences between occupancy prediction algorithms that have near state-of-the-art performance have large effects on energy consumption. We show a 7-9% energy reduction for each 10% decrease in false positive rate and a 13-16% decrease in occupant discomfort for each 10% decrease in false negative rate, demonstrating that even moderate gains in performance could be meaningful in a cost-benefit analysis for a building stakeholder using our tool.

These results apply to the specific building, occupancy traces, and weather data used in our validation. However, they validate the value of our framework as they showcase the utility our framework has for allowing building stakeholders to make informed decisions on: 1) switching from reactive occupancy to predictive occupancy; 2) installing occupancy detection technology and choosing among technology alternatives (switching from a static schedule); and 3) whether to upgrade the prediction occupancy system and algorithms currently in use for the building. A building stakeholder with a substantially different building from our reference implementation, occupancy traces, or weather, could provide the different inputs, and have output that would help to make informed decisions in their own context.

## 2. RELATED WORK

In business, it is common for both the costs and benefits of a proposal to be examined when determining whether to move forward. The results of this analysis come in a number of forms. Most notable is the 'Return-on-Investment' (ROI) time, the time in which it is expected for the proposal to have made back all of the funds spent on its initial outlay. This process exists in almost all forms of business, including building design and modeling.

### 2.1 Building Design and Modeling Tools

There are an enormous number of building design and modeling tools. For the purposes of this paper, we will focus on those that include mechanisms to model the energy consumption of the building or elements of the building. Examples include BLAST [9], DOE-2 [8], ECOTECT [26], EnergyPlus [12], and DesignBuilder [32]. For a recent survey of these tools, see [11].

Building designers and managers, as well as energy analysts, use these tools for a variety of purposes. First, they can be used to ensure that minimal levels of occupant comfort are met by the design of the building, by simulating the temperature and ventilation for each room of the building under a variety of circumstances. Additionally, these tools allow designers to determine the energy costs or savings of various design elements, such as double-paned windows, heavier insulation, and more advanced HVAC systems under various conditions. This allows the designers (and ultimately those buying the building) to make informed cost-benefit decisions regarding which energy savings features of buildings are worth their cost.

Extending the utility of these tools has been a significant focus of prior research. Examples include MLE+, a tool for combining the benefits of MatLab and EnergyPlus for energy-efficient building automation design and analysis, which among other things allows for faster and more building stakeholder friendly analysis [7], and Building Controls Virtual Test Bed (BCVTB), another software tool which acts as middleware for various controllers and Building Simulation systems [36]. As a result, building design and modeling tools have progressed significantly; however, there is still significant work to make these models more accurate.

### 2.2 Occupancy Detection/Prediction

The significant energy consumption of HVAC systems, coupled with a growing worldwide awareness of sustainability, has led to significant research work on methods to reduce their energy usage. Prior work has shown that HVAC control systems can use occupancy data to optimize HVAC scheduling [5][15][17][31]. For example, Erickson *et al.* created an occupancy-based HVAC control system that has the potential to reduce energy consumption by 30% compared to an occupancy-oblivious system, while maintaining thermal comfort [15]. Numerous other predictive techniques achieve similar performance [5][17][31].

One of the key facets of occupancy-based HVAC scheduling is balancing energy cost with occupant "thermal comfort" [10][14][15][16][35]. Keeping an HVAC system on will ensure thermal comfort at all times, but significantly increase energy consumption. Turning off an HVAC system, or running it at minimal levels to keep the building infrastructure intact would achieve a much lower energy usage, but lead to significant occupant discomfort. Neither of these extremes is desirable, which has led to a number of approaches for balancing these factors [5][14][15][16][17][27][31][35].

One of the most promising approaches, reactive control in which a room is adjusted to its preferred temperature only when an occupant is detected, often relies on costly, *per*-office sensors to track occupancy. However, Balaji *et al.* were able to reduce costs by using existing WiFi infrastructure to track participants' smartphones, yielding a 17.8% reduction in HVAC electrical energy consumption [5]. This system moderately reduced thermal comfort during the time taken by the HVAC system to reach the desired temperature of the occupant.

In contrast to reactive occupancy systems, there is potential for modeling occupancy to create a 'predictive occupancy' system [15][22][31]. Scott *et al.'s* work, on the Preheat system for residential homes, utilized predictive occupancy to reduce the amount of 'missed time', time when the house was occupied but not warm, by 6-12 times compared to a static schedule while using an equivalent amount of energy [31].

There has been little attempt to examine the cost-benefit trade of reactive occupancy systems systematically. This is not to say that prior research has been blind to cost. Balaji *et al.'s* work on the Sentinel system was partially motivated to utilize existing infrastructure to reduce installation and running costs [5]. Erickson *et al.'s* POEM system included a ROI analysis [15]. However, neither paper attempts to study the inherent tradeoffs in energy consumption, cost, and comfort systematically.

We present a framework that supports the systematic comparison of occupancy detection/prediction algorithmic performance using several parameters such as accuracy level, error distribution, and building characteristics. Building stakeholders can apply our framework to perform more informed cost-benefit and ROI analyses to evaluate the benefits of installing various occupancy sensing technologies. More information will allow building stakeholders to make better decisions with regards to the aforementioned tradeoff between these benefits and the cost of installation/renovation, and will thus incentive building designers and owners to take advantage of new developments in smart HVAC control with a greater degree of certainty.

## 3. SIMULATION FRAMEWORK

Our framework systematically evaluates the impact of different algorithm parameters on energy consumption and occupant comfort. As shown in Figure 1, our framework uses a multi-step process to simulate the impact of occupancy prediction errors.

At each time step, for each room, the framework determines from the data whether that room will be occupied at a look-ahead time specified as part of the simulation (the look-ahead length). Having determined what an Oracle (perfect predictive system), would do, we then determine the likelihood that a correct prediction will be returned given the provided false positive and false negative rates (see Prediction Step in Figure 1). Importantly, while we ensure that the simulation of occupancy prediction for a given room has the overall provided false positive and false negative rates; we vary these rates throughout the day. It is a well-known fact in building management that occupancy prediction is easier during certain times of day than others. To simulate this effect, we weighted based on the maximum likelihood estimate for a room being occupied or unoccupied at each time of day over the length of our dataset. For example, we use the likelihood of a room being occupied at 3pm each day over the course of the dataset as the weight for the 3pm timeslot. (Note, we calculate weights for each time slot on weekdays and weekends separately, as times that are likely to be occupied during the week are not necessarily likely to be occupied over the weekend).

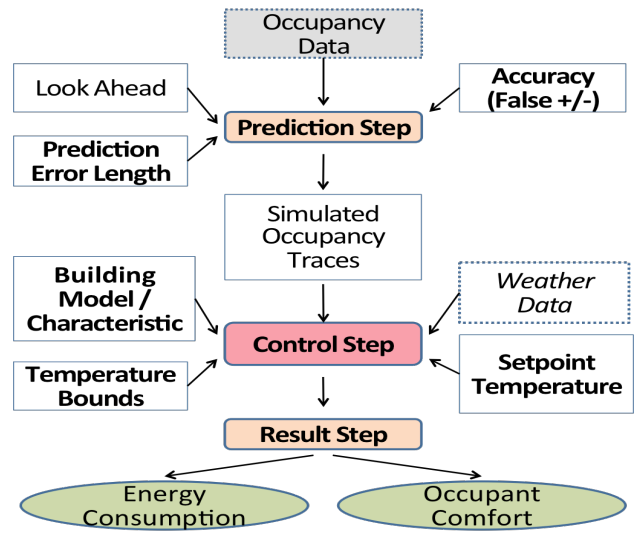

**Figure 1. Three Stages of our Simulation Framework. The control step would involve using some form of existing building modeling software, such as EnergyPlus or DOE-2** *(Dotted boxes: Inputs; Solid boxes: parameters;)*

We redistribute errors so times during which predictions would be more difficult will have higher error rates. Times of the day unlikely to be occupied during our data set will be more likely to have prediction errors occur when they are occupied, as opposed to times of day which are generally occupied, which will be less likely to have errors. Based on this calculated probability we randomly determine whether an error would occur, and then return either the correct prediction, or an error.

Importantly, once the room is actually occupied, we assume that the control system falls back to reactive occupancy, regardless of what the predictive system indicates, and is assumed to correctly monitor the occupant until the occupant leaves the room.

### 3.1 Variations Supported by the Framework

Our simulation framework is designed to support building stakeholder specification and variation along multiple dimensions:

*3.1.1 Accuracy; False Positive and Negative Rate*

The framework can vary accuracy at any level; however, accuracies below 50% are unimportant for analysis, given that the least accurate algorithms in the current literature give significantly higher accuracies [5][16][15][22]. We define accuracy as the chance that if the model were to make a prediction, it will correctly predict occupancy at that level of accuracy. For example, if the correct prediction is a room being 'unoccupied', a model with 70% accuracy would have a 70% chance of predicting 'unoccupied', and a 30% chance of predicting 'occupied'.

Varying accuracy systematically is valuable, but lacks nuance. Most machine learning models have different false positive and false negative rates. Additionally, algorithms can often be tweaked to decrease only one of these error rates, while potentially increasing the other. In light of this, a building stakeholder may wish to examine the effects of differing false positive and false negative rates independently. Importantly, rooms with different occupancy rates will have different overall accuracies at false positive and false negative error rates. A room that is rarely occupied will have fewer errors with a low false positive rate and high false negative rate, than a room that is commonly occupied. To account for this, our framework returns the overall accuracy for the studied rooms to the building stakeholder. This allows the results to be examined by either the false positive/false negative level, or the overall accuracy level, which can be important for predictive occupancy solutions that only report overall accuracies.

*3.1.2 Look Ahead*
A key factor for predictive occupancy control of HVAC systems is how far in advance predictions about occupancy occur. The length of the look-ahead depends both on the temperature change rate of a room as well as the intended deviation from the room's preferred temperature. Too short of a look-ahead may mean not being able to bring the room up to optimal temperatures prior to predicted occupancy, leading to occupant discomfort. Look-aheads that are too long may mean bringing the room up to optimal temperatures somewhat earlier than necessary, leading to HVAC inefficiency. The framework supports a range of potential look-ahead values starting with ten minutes and increasing at five-minute intervals to any value desired. The five-minute interval increase is based on our five-minute data granularity. Changing the interval increase to accommodate more or less fine-grained data is a trivial extension.

*3.1.3 Prediction Error Length*
Prediction models that use temporal features will generally have 'clustered' predictions. In other words, a model that will predict 'unoccupied' at time $t$, is more likely to also predict 'unoccupied' at time $t+1$ and time $t+2$. This is important, because a model that predicts occupied five times and then unoccupied five times, versus one that alternates between predicting occupied and unoccupied, will have differing performance characteristics. To account for this, our model includes a clustering factor when making predictions. While this clustering is an approximation, it is designed to better simulate the patterns of an actual predictive occupancy model using temporal features.

It is important to note that if the model predicts unoccupied for a time length $t$ and the room becomes occupied within this length $t$, our framework will shorten $t$ to account for it. This is in keeping with the earlier stipulation that the system will fall back to reactive occupancy once the room is occupied.

*3.1.4 Temperature Bounds*
HVAC systems are designed to maintain the temperature in a building to levels that prevent damage to the building infrastructure. The temperature in a building is kept well above freezing to prevent water pipes from bursting. Buildings further restrict this range to ensure occupant comfort. However, restricting the range of temperatures for an **unoccupied** room results in wasted energy. Thus, when using our framework, it is important to specify a minimum and maximum allowable temperature, as well as a setpoint for occupied times. These minimum and maximum temperatures can vary depending on the room, to take into account differing heating and cooling preferences for different spaces.

*3.1.5 Occupancy Data*
Occupancy detection takes one of two forms: binary occupancy detection determines whether a room is occupied; and occupancy level detection, seeks to determine the number of people in a given room. Occupancy level detection is still the focus of ongoing research [13], and most buildings use binary occupancy detection. Because of this, we configured our framework to accept binary occupancy data, to improve its general utility.

The base occupancy traces provided, and therefore the simulated occupancy predictions from the prediction step, need not come from the building being modeled. In these cases it is possible to use representative occupancy traces, from a building with similar usage (commercial office building, university building, hospital building) that already has occupancy detection infrastructure, and applying them to the new building. Since the generation of the simulated occupancy traces and the control of the building are entirely separate stages, the fact that the base occupancy traces came from a separate building would have no major effect on the results, assuming that the occupancy traces in question are at least reasonably representative of the target building's usage modalities.

*3.1.6 Building Model*
The characteristics of the building and its HVAC system influence the HVAC system's heating and cooling rate, the rate at which indoor temperature normalizes to the outdoor temperature when the HVAC system is not running, and as a result the overall energy use of the building. Many factors affect the performance of an HVAC system, including the outdoor temperature, cloud cover (as the effect of solar energy on a building's internal temperature can be significant), insulation, HVAC system design, and the current efficiency of HVAC components.

Due to the number and complexity of these factors, perfectly modeling HVAC heating and cooling rates, temperature normalization or decay rate, and overall energy consumption is a difficult problem. As a gold standard, we use Revit buildings models with Energy Plus simulations to model temperature change and overall energy consumption [4][12].

## 3.2 Supported Metrics
*3.2.1 Energy Impact*
As mentioned previously, we use a Revit model and Energy Plus to simulate a building's HVAC control in response to our simulated occupancy predictions. EnergyPlus calculates HVAC energy usage on a monthly basis at a room-by-room level, and we use this calculation as our Energy Impact metric. We aggregate the room-by-room and monthly data to analyze at the building level over the entire simulation period for each simulation.

As stated above, a building stakeholder can choose to use another simulation tool, or even utilize a custom energy calculation that does not depend on a building model-based technique. In this case, we recommend an equation provided by the U.S. Department of Energy (DoE) [33], which has been used in the research literature [6]. While inherently less accurate than a building model simulation, this equation is generally applicable and available to all building stakeholders.

*3.2.2 Occupant Discomfort*
Our framework supports an estimate of occupant discomfort by analyzing MissTime [25], the average daily number of minutes a room was not at the correct comfort temperature. This metric indicates on average how many minutes per day an occupant had to endure temperatures different from her preferred temperature. We differ from Preheat, an accurate occupancy prediction algorithm that used MissTime as a measure of occupant discomfort [25] insofar as we assume a fixed preferred comfort temperature and not a comfort temperature range. This limitation is necessary since EnergyPlus does not give us the ability to export the daily room temperature distribution in a 5-minute interval and thus we cannot calculate when a room reaches the comfort temperature range. This will overestimate the comfort impact, but we believe it will still provide a valuable validation point for our framework.

We acknowledge that other factors, such as metabolic rate, clothing level, and humidity effect comfort; however, this information is highly individualistic and generally unavailable to building managers. Therefore, we argue that a simple temperature based comfort metric has greater utility for our framework.

# 4. FRAMEWORK VALIDATION

The goal of our validation is to show specific differences in the performance of simulations varying algorithm parameters that building mangers could use to make better decisions as to what type of occupancy prediction hardware and software will be most appropriate for their needs.

Unfortunately, the Revit model for the building that we used to collect occupancy traces was unavailable. Instead, we used the building model for another representative building and created a mapping between the two buildings. While not ideal, this does not significantly reduce the validity of our results. The simulation of occupancy prediction and control of the building are separate stages of our framework. Additionally, this validation showcases that varying the factors we have chosen to analyze has a an effect on a building's energy consumption and occupant comfort. We do not seek to show that this is true for a specific building, rather to provide a proof of concept that this framework yields results that can be useful for real buildings generally. Since our Revit model accurately depicts an existing building, its results will serve just as well to validate our framework.

Currently, the building uses a reactive occupancy system. We use this control system as a baseline to compare our simulations against. Additionally, we simulated a static 6am-9pm schedule as another baseline to compare against.

## 4.1 Simulation Parameters

Our simulation framework has a wide range of inputs and parameters. We focus on varying the most important parameters in terms of effect on our metrics: accuracy and temperature bounds, for a total of 27 simulations. Additionally, we simulate a reactive system and a static schedule system, both as baselines to compare against, as well commonly used control systems.

*4.1.1 False Positives/Negatives: Many variations*

The primary goal of our validation was to showcase simulating different levels of false positive and false negative rates, and compare their impact given other parameterizations of the framework. We chose to focus our analysis on false positive and false negative rates as most designs of predictive occupancy algorithms and control systems report their performance either in terms of overall accuracy or false positive and false negative rates. Our experiments were conducted at 25%, 15%, and 5% false positive rates crossed with false negatives rates. (9 total error rate variances). We chose these values as representative of the range of performance for 'state of the art' prediction algorithms.

*4.1.2 Temperature Bounds: Three Variations*

Setting temperature bounds relies on two variables: the heating and cooling setpoints when a room is occupied, and the length of the bounds between occupied and unoccupied setpoints. We chose 20°C (68° F) as our occupied heating setpoint, and 24°C (75°F) as our occupied cooling setpoint. We chose these setpoints based on the ISO 7730 standards of comfort [21]. We also simulated three different temperature bounds, as described in Table 1: The first bounds (*Small-Bound* in Table 1) are 2°C from the setpoint temperatures. We chose these as the most rigid realistic bounds, similar to those used in reactive control systems. The second set of bounds (*Medium-Bound* in Table 1) are 6°C from the setpoint temperatures. We chose these as they were within safe operating parameters for the building, but would provide some notable energy savings, without jeopardizing occupant comfort. Finally, the last set of bounds (*Large-Bounds* in Table 1) is 12°C from the setpoint temperatures. These represent the most extreme temperature bounds that a building manager could use, as allowing the temperature to float further could have negative effects on building operation. This final condition is used to show the upper bound on energy gains that a building stakeholder could achieve. Additionally, we run the reactive system with small-bounds, and the static schedule with the large-bounds case as comparisons.

**Table 1: Simulation Conditions**

|  | Predictive Occupancy | Reactive Baseline | Static Baseline |
|---|---|---|---|
| **Temperature deviation bounded by** 2°C (~4°F) **setback** | Run | Run |  |
| **Temperature deviation bounded by** 6°C (~11°F) **setback** | Run |  |  |
| **Temperature Deviation bounded by** 10°C (~20°F) **setback** | Run |  | Run |

Our building strongly restricts the extent to which occupants can control their own rooms temperature (our test building was limited to 4°F), and only during occupied times, limiting the utility of individual occupant specific temperature bounds. Thus, to reduce complexity, we used a standard set of temperature bounds

*4.1.3 Fixed Values*

To focus our analysis on our pivotal framework parameters, we chose fixed values for our prediction error and look ahead lengths. We configured our framework to use *Random-Prediction* errors length, lasting randomly from five minutes (the shortest time-step in our dataset) to one hour. These random prediction errors had a mean length of 32.5 minutes (SD=13.8 minutes). We chose 60 minutes for our look-ahead length, both as a reasonable time frame to perform predictions over, as well as matching our longest possible prediction error length.

*4.1.4 Occupancy Traces*

Our data set includes 196 days of occupancy data collected at five-minute intervals, from June 2011 to December 2011 for 235 rooms in our test building. This data included a timestamp for when the data was collected and a binary occupancy status. Occupancy was measured with commonly used dual-technology (Passive Infrared and Ultrasonic) sensors deployed in our testbed building, networked to the existing Building Management System.

*4.1.5 Building Modeling:*

We use a Revit Building Model to perform EnergyPlus simulations of energy consumption caused by our simulated predicted occupancy traces. Our simulation included Pittsburgh weather data over our occupancy collection period. The multi-seasonal nature of our data collection period serves to our framework over a range of weather periods.

It is important to note that we did not simulate the entire building,

only rooms for which we had collected occupancy traces. This has little bearing on comparisons between conditions however.

### 4.1.6 Evaluation Metrics

To quantify the energy impact of different options we use the output of our EnergyPlus simulations. For occupant comfort, we use the aforementioned variant of MissTime. We compare in terms of both absolute, and percentage reductions.

## 5. EVALUATION RESULTS

For each simulation scenario we generated, we calculated two metrics (totaled over the 6 month period of our occupancy traces) – Energy Impact and Occupant Comfort. We compare the predictive occupancy simulations to themselves, to the reactive baseline, and to the static schedule baseline.

### 5.1 Energy Impact

Figure 2 shows the heating and cooling HVAC energy usage for the offices we simulate in our test building under each of the predictive occupancy simulation conditions in kWh.

As expected, wider temperature bounds reduced energy impact in each case. While unsurprising on a theoretical level, it is important to note that rather than having a theoretical understanding that wider temperature bounds reduce energy usage, these results let our building manager know specifically that widening from the small to medium bounds led to an average overall savings of 149,000 (kWh), or 24%, and between the medium temperature bounds and large temperature bounds and average overall savings of 124,379.84 (kWh) or 28%. This is particularly important as the relationship between widening the temperature bounds (by number of degrees) and energy reduction is obviously non-linear, and thus not easily calculated without testing specific temperature bounds.

The second important result in Figure 2 is also one that is theoretically expected. Decreasing the simulated false positive rate led to declines in energy usage under all temperature bounds. However the value of our framework comes from quantifying this expected result in useful terms. A building stakeholder would know that a decrease from 25% to 15% false positive rate showed an average 18,000 kWh reduction in energy impact, or 7.27% total power usage, and similarly that a decrease from 15% to 5% false positive rate showed an average 22,000 kWh reduction in energy impact, or 9.30% total power usage. This is a particularly useful piece of information because the relationship between the false positive rate and energy impact is non-linear, and so the increased number of datapoints our framework provides is valuable for better representing that non-linearity.

### 5.2 Percent Savings Compared to Baselines

Figure 3 shows the relationship between overall energy consumed by each of the 9 simulated occupancies, and the reactive and static baselines established. As expected, the reactive system outperforms all levels of predictive occupancy using the same small temperature bound. However, it is important to note that the predictive medium and large bound cases both outperform the reactive system. More importantly, our framework quantifies how much these larger bound cases outperformed the reactive system. Depending on the false positive and false negative rates, our building stakeholder could expect to see a 10.68% to a 26.62% reduction in energy usage by changing from a reactive system with small temperature bounds to a predictive occupancy system with medium temperature bounds. Similarly, the conversion to a large temperature bound system would yield a 16.44% to 40.73% reduction in energy usage.

The static 6am-9pm schedule with large temperature bound outperformed all predictive occupancy cases with small temperature bounds. However, the occupancy cases with large and medium temperature bounds both uniformly outperformed the static system. This is somewhat unsurprising in the case of the large temperature bounds, since most occupancy occurs during the day and therefore the static system will be 'on' more often than the predictive system. However, it is important to note that our framework provides building stakeholders with quantified values for these gains, from 14.7% to 39.5%. The fact that the predictive occupancy cases with medium temperature bounds also outperform the static schedule is more surprising, since the effect of temperature bounds is generally quite large, and would therefore not necessarily be intuitive to a building manger. In addition, our framework quantifies the amount that the predictive occupancy algorithms with medium setbacks outperform the static system by, showing an energy impact decrease from 8.8% to 25%.

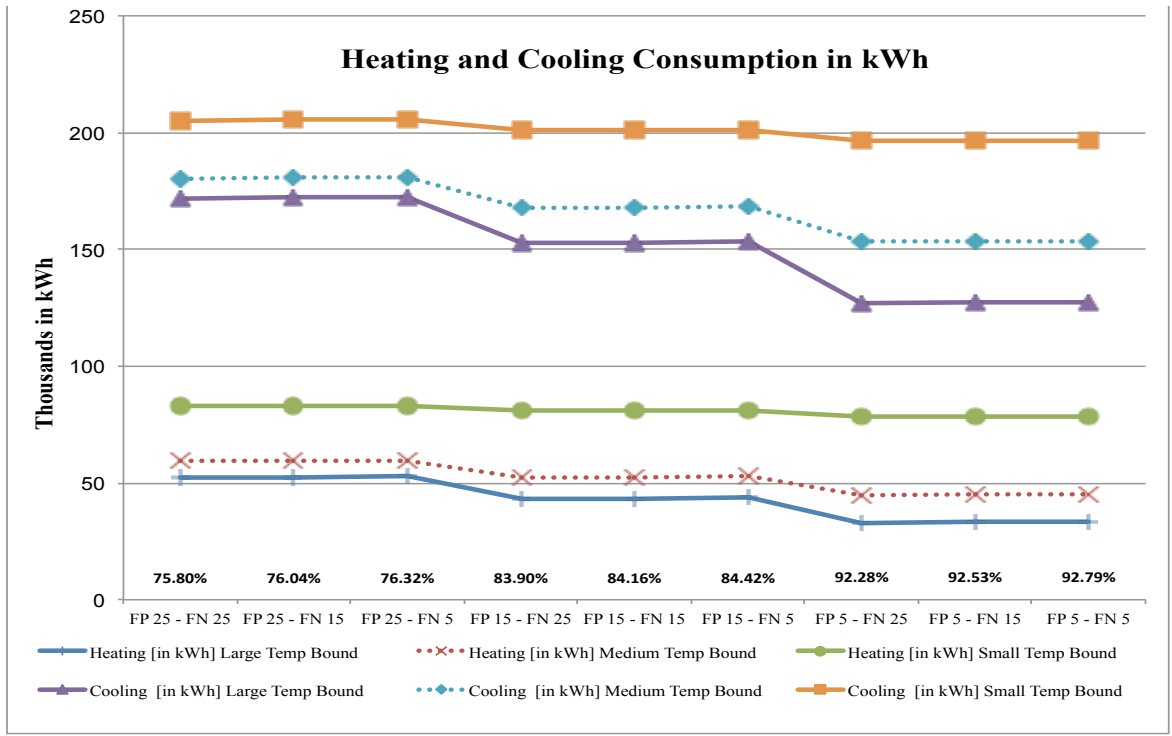

**Figure 2 Overview showing energy consumption, in kilowatt-hours, of the 27 Energy Plus predictive occupancy simulations (3 FP rates x 3 FN rates x 3 temperature bounds) run over 6 months. % for each datapoint is average accuracy across all rooms.**

## 5.3 Occupant Discomfort

We also report the results of our occupant discomfort metric, as can be seen in Figure 4. As expected, the lower false negative rates had significant impacts on the average daily MissTime. However, once again our framework allows us to quantify this effect for our building stakeholder: the average difference between the 25% false negative rate and 15% false negative rate was 5.3 minutes a day, or 13.5%, and the average difference between the 15% false negative rate and 5% false negative rate was 5.1 minutes a day, or 15.6%. While comfort does not lend itself as well to a direct price comparison, the quantified difference in occupant comfort that different predictive algorithms would achieve has obvious utility for building stakeholders.

Unsurprisingly the predictive occupancy systems outperformed the reactive system in terms average daily MissTime. The reactive system in fact had one of the worst MissTime at 213 average daily minutes; however, this may be due to the strict standard of comfort that we describe in Section 3.2.2. Nevertheless, our validation provides our building stakeholder with quantified values of this reduction: 170-185 average MissTime minutes a day less in predictive occupancy cases.

Finally, the predictive occupancy systems also outperform the static schedule in terms of occupant comfort. This result is somewhat surprising, since the static schedule has perfect occupant comfort from 6am-9pm. Nevertheless, the static system has 124 average missed time minutes, suggesting that on average there is about 2 hours of occupancy a day outside of this time band. This is between 80-95 average more MissTime minutes than the predictive occupancy systems.

It is important to note that these results may be specific to university occupancy patterns, with students and faculty working odd hours. While they are important for our building manager, a building stakeholder in an industry office building might need to use occupancy traces either from their building, or from a similar industry office building, to achieve more accurate results.

## 6. DISCUSSION

The value of our results comes from showing the type and scope of information available to building stakeholders to make informed cost-benefit analyses. Therefore, we discuss the effects that our results might have on the decision making process of a building stakeholder of our building in a number of cases, understanding that a building stakeholder using our framework for a different building would have the same information, if not the same exact results, to inform their decision.

Note that the specific results presented above apply only to our test building in our university's climate with the occupancy traces we recorded over a 6-month period. This means that our results do not show that certain occupancy detection and inference algorithmic parameters are better for all buildings, climates, and occupancy traces, just for our specific building.

### 6.1 Reactive vs. Predictive Occupancy

As a point of common knowledge supported by our validation, a reactive occupancy control system outperforms predictive occupancy control systems with regards to energy consumption when the same temperature bounds are used for our building. However, as stated at the beginning of the paper, to ensure occupant comfort, reactive systems only use small temperature bounds, whereas predictive control systems can use much larger temperature bounds. In the case of our building in our validation, the medium temperature bound predictive occupancy systems outperformed the reactive system in terms of energy consumption, while maintaining occupant comfort. These predictive occupancy cases consumed approximately 20.8% less energy than the reactive case. A building stakeholder for our building could therefore expect to achieve a 20.8% energy savings, depending on the predictive occupancy algorithm she deployed, while maintaining occupant comfort.

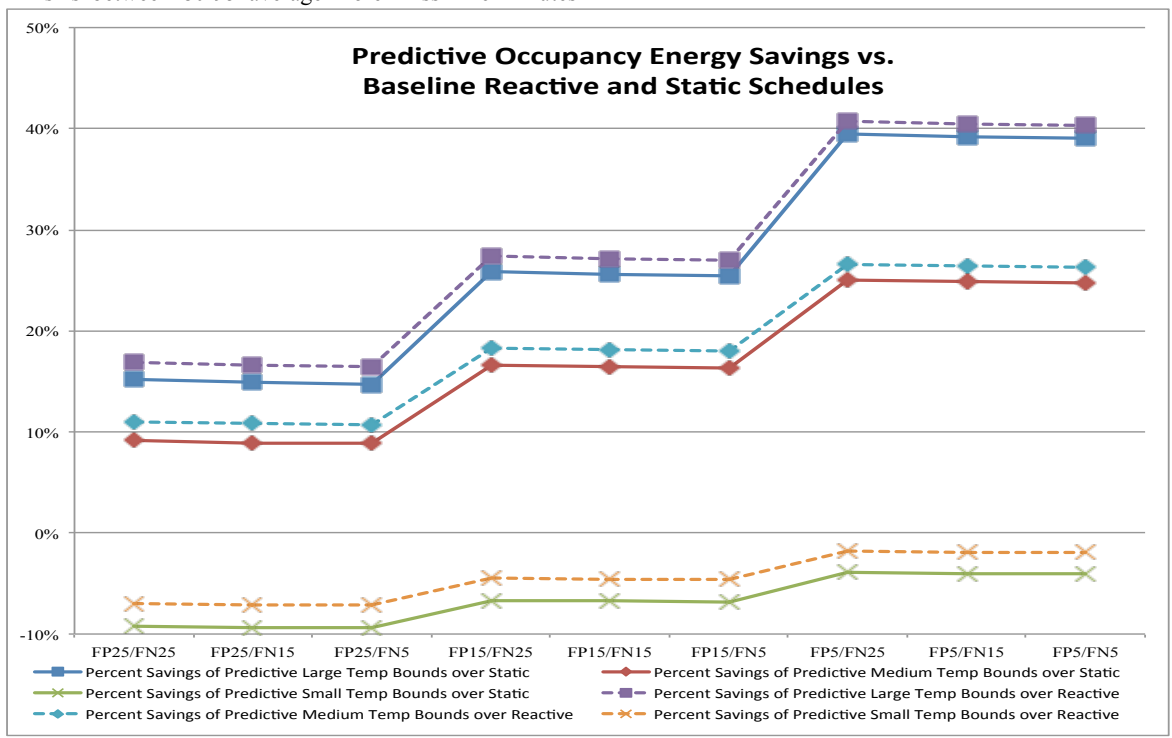

**Figure 3. Energy Savings of Predictive Occupancy systems vs. Baseline Reactive and Static Schedules.**

Depending on the costs of electricity and predictive occupancy software or hardware, for algorithms requiring more complex sensing technology and feature sets [15] our building stakeholder could confidently choose a software package or hardware upgrade that maximizes her savings, with the added benefit of reducing overall energy usage for sustainability purposes.

## 6.2 Static vs. Predictive Occupancy

The discussion above focuses on a building stakeholder who already has occupancy detection technology installed, and is considering the best way to employ or upgrade it. Given the large proportion of buildings still using static control schedules, there are many building stakeholders who do not have occupancy detection technology installed, and have too little information to make informed decisions as to whether to install occupancy detection technology, as well as how to use it. Our framework could help inform these decisions.

We ran a single simulation with a static control schedule, which kept the temperature to the setpoints between 6am and 9pm, and allowed the temperature to float to the large temperature bounds at all other times. As a comparison, the medium temperature bounds predictive occupancy simulations performed significantly better than the static schedule in terms of energy consumption for our building, consuming approximately 17% less energy than the static system. Additionally, the results for occupant comfort show that the predictive occupancy systems significantly outperformed even a 15-hour static schedule for our building. We note that this occupant comfort result may be an artifact of using a university building, with wider hours, than a general office building.

Our test building had occupancy detection hardware, allowing us to collect occupancy traces for our validation. However, building stakeholders trying to determine if they want to change from a static control system to a predictive occupancy control system could still inform their decision to potentially install more complex (*i.e.*, more expensive to install and maintain) control systems based on our framework's results, the cost of said control systems, and the price of electricity.

Thus far, we have focused on buildings with occupancy detection infrastructure already in place. This is important, as occupancy traces are a necessary input for our framework. In the event that a building does not have occupancy detection technologies in place, assuming that building stakeholders had their planned building model and some climate data for the area they plan to build or have already built their building, then representative occupancy traces from another building could be used to inform decisions as to whether to install occupancy detection systems. Taking occupancy traces from another building would almost certainly lead to some uncertainty in results; however, utilizing occupancy traces for similar buildings could minimize this.

## 6.3 Comparing Predictive Occupancies

Another important case is considering whether to upgrade a pre-existing predictive occupancy system in a building. As has been mentioned many times in this paper, the field of predictive occupancy is a vibrant one, and new predictive solutions and systems come out every year. Therefore, it is quite possible that a building stakeholder using one predictive occupancy control system might want to consider a new predictive solution.

Our results illustrate the benefits of various levels of predictive occupancy algorithmic performance for our test building. Our building stakeholder, assuming they were running a predictive system with a 15% false positive rate would know that, depending on the temperature bounds they decided to employ, they could achieve 5-10% energy savings. This would allow our building stakeholder to make an informed decision balancing energy costs with the cost of deploying a new control system.

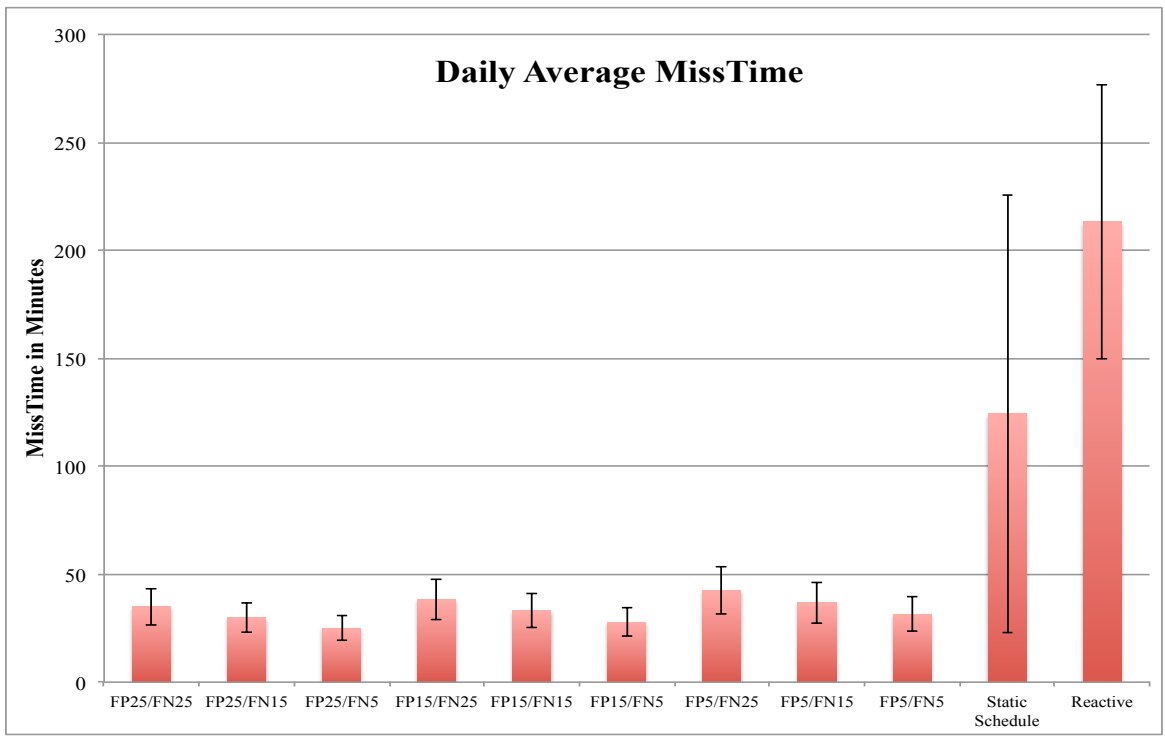

**Figure 4 Average Daily Occupant Discomfort across predictive occupancy conditions and baselines. Error bars denote standard deviation.**

While the specific 5-10% energy savings apply only to the building we study, any building stakeholder with predictive occupancy solutions already installed could utilize our framework and then use results specific to their building to inform their own decision as to changing or upgrading systems.

## 6.4 Generalizability

The results from the validation of our framework are based on the occupancy traces and building model we used, in the climate our university is located in. However, the analytic framework we developed has significant broader applicability. An analysis of the occupancy for our validation shows a mean occupancy of 20.2% (SD=11.02%). The high standard deviation demonstrates that rooms had varying occupancy rates, and showcases both that current building-wide static schedules are not efficient in such a building, and the data upon which we base our validation represented a range of occupancy patterns. The results from our validation remain specific to our building and occupancy traces; however, this discussion illustrates that our data is not some form of edge case for occupancy traces/buildings/climate, and that we tested our framework on a fairly representative building. Additionally, our framework can be easily applied to other settings by providing new occupancy traces, building models, weather data and details of the predictive system being considered.

## 6.5 Utility

The proliferation of building simulations tools allowing the calculation of a building's energy consumption show the value of such simulation in cost-benefit analyses. We expand this utility by including the occupancy detection and inference system accuracy, in terms of false positive and false negative rates, as factors that can be incorporated into the cost-benefit analysis. Additionally, while algorithmic accuracy is the most commonly reported metric for predictive occupancy solutions, a building stakeholder could also simultaneously compare the effects of multiple other factors, for example those related to the prediction algorithm, such as prediction error length and look-ahead length, and those related to building management, such as the temperature bounds to set. Also, for building stakeholders who already have detection and predictive technology in place, they may choose to seek out and make use of more or less complex and or expensive control algorithms (*i.e.*, software) based on the benefits that our framework estimates such algorithms would achieve.

Our validation varied control system factors, such as false positive and false negative rates; however, building stakeholders could also use our framework to make decisions for potential design elements for future buildings (e.g. insulation type) or renovations.

## 6.6 Limitations & Future Work

The results of our analysis validate the utility of our framework; however, it is important to put them in context. While Revit models and EnergyPlus simulations represent a gold standard for building modeling, any model has limitations. In particular, determining the errors bounds of Energy Plus simulations, to provide building stakeholders with a quantified value of the potential inaccuracies of our results, remains a matter for future work. While the effect of such error bounds will be mitigated by the fact that we perform within-building comparisons, they may have some impact on the results

Our framework currently requires a building stakeholder to have representative occupancy data, outdoor temperature data for her building, and some form of building model at a minimum. Temperature data, even at a somewhat fine-grained level, is relatively easy to find, and any building model that takes occupied time as an input can be used. However, a building stakeholder may not have access to representative occupancy traces. Therefore, one target for future work will be to create a repository of occupancy traces, collected by us or other researchers [2], for buildings of different general use types. Building stakeholders could then choose the occupancy trace that most closely mirrors their own, to get a good approximation of performance.

Finally, our current occupant comfort metric, while useful, is somewhat coarse-grained. Developing a better metric, possibly achieved through changes to the EnergyPlus source code or API, is an important avenue for future work. Our main goal would be to calculate comfort based on a variety of factors, including the degree of temperature difference and occupant specific factors.

## 7. CONCLUSION

We have presented the design, implementation, and validation of a framework supporting building stakeholders in making informed decisions regarding their method of occupancy detection and inference and HVAC control. Our framework allows building stakeholders to explore and quantify the effect of a wide range of factors, such as predictive accuracy, temperature bounds, prediction error length, and building model, in terms of both energy consumption and occupant comfort. Our results show that analyzing the effects of varying occupancy prediction strategies yields quantified information, on the benefits of certain simulated algorithms, which any energy analyst, architect, or building manager, could use fairly easily.